\documentstyle[11pt]{article}
\newcommand{\doublespace}{
    \renewcommand{\baselinestretch}{1.6}\large\normalsize}

\newcommand{\be}{\begin{equation}}
\newcommand{\ee}{\end{equation}}
\newcommand{\ba}{\begin{eqnarray}}
\newcommand{\ea}{\end{eqnarray}}

\newcommand{\ave}[1]{\langle {#1} \rangle}

\def\roughly#1{\mathrel{\raise.3ex\hbox{$#1$\kern-.75em%
\lower1ex\hbox{$\sim$}}}}
\def\fm{{\rm fm}}
\def\lsim{\roughly<}
\def\gsim{\roughly>}
\def\ds{\displaystyle}
\def\MeV{{\rm MeV}}
\def\fm{{\rm fm}}
\def\pl{Phys. Lett.}
\def\np{Nucl. Phys.}
\def\pr{Phys. Rev.}
\begin{document}
\begin{titlepage}
\pagestyle{empty}
\vspace{1.0in}
\begin{flushright}
SUNY-NTG-96-6
\end{flushright}
\begin{flushright}
March 1996
\end{flushright}
\vspace{1.0in}
\begin{center}
\doublespace
\begin{Large}
{\bf{A MEAN FIELD THEORY}}\\
{\bf OF THE CHIRAL PHASE TRANSITION}\\
\end{Large}
\vskip 1.0in
G.E. Brown, M. Buballa\\
{\small
{\it Department of Physics, State University of New York,\\
Stony Brook, New York 11794, U.S.A.}}\\
and\\
Mannque Rho\\
{\small
{\it C.E.A.-- Saclay, Service de Physique Th\'eorique\\
F-91191 Gif-sur-Yvette Cedex, France}}
\end{center}
\vspace{2cm}

\begin{abstract}
The recent discussions by Koci\'c and Kogut on the nature of the chiral
phase transition are reviewed. The mean-field nature of the transition
suggested by these authors is supported in random matrix theory
by Verbaarschot and Jackson which reproduces many aspects of QCD lattice 
simulations. In this paper, 
we point out physical arguments that favor a mean-field transition, not only
for zero density and high temperature, but also for finite density. We show,
using the Gross-Neveu model in 3 spatial dimensions in mean-field 
approximation, how the phase transition is constructed. 
In order to reproduce the lowering of the $\rho=0$, $T=0$ vacuum evaluated
in lattice calculations, we introduce {nucleons} rather than constituent
quarks in negative energy
states, down to a momentum cut-off of $\Lambda$.  
We also discuss Brown-Rho
scaling of the hadron masses in relation to the QCD phase transition, and
how this scaling affects the CERES and HELIOS-3 dilepton experiments.  
\end{abstract}
\end{titlepage}

\doublespace

\section{Introduction}

\setcounter{equation}{0}

Until recently, the finite-temperature chiral restoration transition in QCD
has been guided by the concepts of dimensional reduction and universality.
Pisarski and Wilczek \cite{PW84} and Wilczek \cite{W92} described the QCD
chiral transition with two light quarks near the chiral transition by the
(dimensionally reduced)
three-dimensional $\sigma$-model. The fluctuations of the order parameter,
$\sigma$ and $\pi$, go soft at the transition temperature. Since $\sigma$
and $\pi$ are bosons, they have zero modes $\omega_0 = 0$ in their finite
temperature Matsubara decomposition. These zero modes dominate the infrared
behavior, and, therefore, the phase transition. The consequence is that the
chiral transition of four-dimensional QCD should lie in the same universality
class as the three-dimensional $O(4)$ magnet \cite{PW84,W92}.  

Koci\'c and Kogut \cite{KK95,KK95P} recently pointed out that the fermionic
nature of the quark and antiquark, the constituents of the $\sigma$-meson,
can fundamentally affect the nature of the phase transition. This is easily
seen by noting that the sum over Matsubara frequencies in the 
finite-temperature fermion Green's function brings in the factor
\be  
F = [ \frac{1}{2} - \frac{1}{e^{\omega/T} + 1} ] \; , 
\ee
which vanishes for $\omega \rightarrow 0$. The $\omega = 0$ states of the
fermions in the heat bath at finite $T$ are all filled, so it is not
possible to introduce an additional $\omega = 0$ fermion into the system.
Similar
arguments lead Koci\'c and Kogut to suggest that the QCD chiral restoration
transition could be mean-field in nature.

The arguments based on the thermal factor eq.~(1.1) would also hold for
finite density, except at $T = 0$. Whereas relatively firm arguments 
about the nature of the phase transition can be made only for zero density
and finite $T$, we shall assume that the transition is of mean-field type
for finite
densities and temperatures. This will be in line with our use of a scalar
mean field as order parameter. 

Recently, Verbaarschot \cite{V95P} has shown that the Columbia valence quark
mass dependence of the chiral condensate evaluated in QCD lattice simulations
with dynamical fermions for a series of couplings close to the critical
temperature can be well described by chiral random matrix theory. The
randomly drawn matrix elements of the Dirac operator are constants,
independent of space or time. The supposition can be made that the chiral
restoration transition is governed by the mean field, also independent of
space and time. Jackson and Verbaarschot \cite{JV95P} have formulated a
random matrix model which mimics the chiral phase transition with two light
flavors. They find mean-field values for the critical exponents $\beta$ and
$\delta$.
It could well be that movement towards $T_c$ is described well by mean field
until $T$ is very close to $T_c$, and then changes character. In the case of
BCS superconductivity, the ``Ginzburg window" for the phase 
transition \cite{Ginz} is $t\sim 10^{-14}$ where $t = \frac{T_c-T}{T_c}$;
down to this window the transition is well
described by mean field. The window for something like the Pisarski-Wilczek
behavior \cite{PW84,W92} to be realized is small, according to our above
arguments, so the mean-field assumption will be adequate for us.

As noted by Koci\'c and Kogut \cite{KK95,KK95P} and in earlier references
in these works, the Nambu--Jona-Lasinio model has at finite temperature a
chiral restoring transition in mean field (with mean-field critical exponents).
They point out \cite{KK95P} that one expects meson masses to gradually decrease
to zero, and meson radii to increase, with movement in temperature towards
symmetry restoration.  Furthermore, lattice calculations \cite{G91} find 
that the screening masses of the $\rho$
and $a_1$ mesons go to $2\pi T$ as $T \rightarrow T_c$, whereas they would be
\be m_{scr} = \sqrt{(2\pi T)^2 + m_{dyn}^2}  \ee
were a dynamical mass $m_{dyn}$ left at $T_c$ \cite{BR95P}. It turns out
that to the accuracy
of the lattice calculations, no $m_{dyn}$ is needed.

\section{Walecka Mean Field Theory}

\setcounter{equation}{0}

Walecka mean-field theory describes well a number of nuclear phenomena. We
take this to mean that it works well at nuclear matter density. 
Several recent developments support this assertion: (1) Gelmini and
Ritzi \cite{GR95P} proposed deriving Walecka mean-field theory from chiral
mean-field theory, treating both vector and scalar fields as chiral singlets;
(2) Brown and Rho \cite{BR95PP} showed that Brown-Rho (BR) 
scaling \cite{BR91} played
a key role in obtaining the Walecka parameters at nuclear matter density
$\rho = \rho_o$; (3) it has been shown by Park, Min and Rho \cite{pmr95}
that BR scaling can be derived in chiral perturbation theory in medium
with the inclusion of multi-Fermi interactions in the chiral Lagrangian
and (4) it has been argued by Friman and Rho \cite{friman} that the mean-field
Lagrangian theory with BR 
scaling can be mapped to the Landau Fermi liquid fixed point,
with the Landau parameters completely describing low-energy spectroscopic
properties of heavy nuclei.

In order to describe density-dependent constituent quark masses in the 
analysis of dilepton production in the CERN SPS relativistic heavy ion
reactions, Li, Ko and Brown \cite{LK95P} employed Walecka theory at constituent
quark level in the transport equations; i.e., in the relativistic VUU
\cite{KL87} equations. A satisfactory treatment, which is thermodynamically
consistent (conserving energy, etc.), can be obtained in this way.

Although, as explained later, the Walecka mean-field theory cannot be carried
blindly
all of the way to the chiral restoration transition, we can still gain a
valuable insight from it into how chiral restoration is approached.
First of all, we remark that the coupling constants of the Walecka linear
sigma-omega model, chosen to fit nuclear phenomena, are satisfactory for
the constituent quark model. The $g_{\sigma NN} \simeq 10$ of the linear 
sigma-omega model \cite{SW86} becomes
\be 
g_{\sigma QQ} = g_{\sigma NN}/3 \cong 10/3    
\ee
for the constituent quark $Q$. The mean-field Goldberger-Treiman relation is
\be 
m_Q = g_{\sigma QQ} f_\pi   \;,     
\ee
with $g_A^Q=1$ (implied by the large $N_c$ argument) which gives
\be 
m_Q = 310\ {\MeV}    
\ee
using eq.~(2.2) and $f_\pi = 93$ \ MeV. This seems satisfactory in that $m_Q$
is about $1/3$ of the nucleon mass. We therefore propose to treat 
the light hadronic degrees of freedom in terms of light-quark constituent
quark degrees of freedom. In doing this confinement is not considered 
explicitly. When necessary, we will put the confinement effect by hand
as will be the case with the calculation of the bag constant. 

We now show that the scalar field energy in Walecka theory acts like a bag
constant of the hadronic sector as $T \rightarrow T_c$. In other words
\be 
\varepsilon_{Field} = \frac{1}{2} m_\sigma^2 \sigma_W^2 = B_{eff} \;, 
\ee
where $\sigma_W$ is the Walecka scalar mean field. That $\varepsilon_{Field}$
behaves like a bag constant can be seen from the expression for the pressure
\be 
p = n_B \frac{d \varepsilon}{d n_B} - \varepsilon \;, 
\ee
where $\varepsilon$ is the energy density and $n_B$ the baryon number.
Since the field energy is independent of $n_B$, $p = - \varepsilon_{Field}$, 
which is just the behavior of the bag constant $B$.

Neglecting the fact that in Walecka theory, the effective mass $m_Q^*$ goes
to zero only as the density $\rho \rightarrow \infty$ (we shall repair this
defect in the next section), let us evaluate eq.~(2.4) for $m_Q^*  \rightarrow
0$:
\be
B_{eff} = \varepsilon_{Field}(m_Q^* = 0) = \frac{1}{2} m_\sigma^2 \, M_N^2 / 
g_{\sigma NN}^2 \simeq \frac{1}{2} m_\sigma^2 \, m_Q^2 / g_{\sigma QQ}^2 
= \frac{1}{2} m_\sigma^2 \, f_\pi^2  \;,
\ee
where we have used $M_N^* = M_N - g_{\sigma NN} \sigma_W$ from Walecka theory
and eqs.~(2.2) and (2.4). 
Now Walecka chooses $m_\sigma = 550$ \ MeV. In the interpretation of Brown and
Rho \cite{BR95P}, this $m_\sigma$ should be $m^*_\sigma(\rho_o)$, and the
$m_\sigma$ at zero density would be, according to their arguments,
$m_\sigma \cong m_\sigma^*(\rho_o) / 0.78 \cong 705$ \ MeV. The factor
0.78 is the 
computed ratio $f_\pi^*(\rho_o)/f_\pi$ and in Brown-Rho scaling \cite{BR91},
the scalar masses scale as $f_\pi$. With this $m_\sigma$
\be
   B_{eff} \;=\; 280\ {\MeV} / {\rm fm}^3 \;.\label{B}
\ee
This is somewhat uncertain in magnitude, because there is no real scalar meson
of mass $m_\sigma$; the value $705 \ \MeV$ (or the
\underline{in}-\underline{medium} $550 \ \MeV$ used in nuclear calculations) 
is an
effective value, used in mean-field calculations, to describe an enhancement
in the $S = 0$ two-pion-exchange. This may be interpreted as the
``dilaton" field that
joins the pion field at the phase transition as described in \cite{BR95P}.
The value (\ref{B}) may be a bit high,
because Walecka mean-field calculations are mostly performed in nuclei at a 
density $\rho < \rho_o$, so that the extrapolation $\rho = 0$ will not bring in
quite as large an $m_\sigma(\rho=0)$. Corrections for such effects are within
our uncertainties. 

We suggest that it was no coincidence that Koch and Brown \cite{KB93} found
the change in quark and gluon condensate, as $T$ went upwards through the
phase transition at $T_c$
\be
\delta B \sim \frac{1}{2} B \;,
\ee
where $B$ is the full gluon condensate \cite{S78}
\be
 B \;=\; 471\ {\MeV}/{\rm fm}^3  \;=\; (254\  {\MeV})^4 \; .
\ee
$\delta B$ is the bag constant which should be used in order to determine the
pressure in the {\it quark/gluon phase}; 
i.e., for massless noninteracting quarks and gluons
\be
p = 37 \frac{\pi^2}{90} T^4 - \delta B \;.
\ee
The $\delta B \sim 250\  {\MeV}/{\fm}^3$ found by Koch and Brown \cite{KB93}
is only $\lsim$ half of the full gluon condensate
\cite{S78}. Whereas it is known that only part of the glue melts at the
chiral restoration transition (see \cite{AB93} for a full discussion of this),
it has been somewhat of a mystery why just about half of the condensate melts.

We shall show later that, in order to construct a second-order chiral restoring
transition, $B_{eff}$ on the hadron side must equal the $\delta B$ on the
quark-gluon side of the phase transition. In other words, in the hadron sector,
the vacuum energy must be raised just enough to make the quarks massless, and
then the transition can take place. We believe that this determines the
necessary fraction of the gluons which have to melt before the phase transition
can take place. In fact, the increase in vacuum energy must be given generally
as a contribution to $T^{oo}$ of the stress-energy tensor. {\it 
This contribution is
expressed below $T_c$ in terms of hadronic degrees of freedom, and above $T_c$
in quark/gluon ones}.
It is, thus,
not surprising that the bag constants at two sides of the phase transition
are equal at $T_c$, because they are the same $T^{oo}$, expressed in
different variables.

\section{Review of the Chiral Restoring Transition in Walecka Theory}

\setcounter{equation}{0}

The chiral restoring transition has been constructed in Walecka theory
\cite{TG83}. Even though the Walecka theory is not manifestly chiral
(although it can be made so at mean-field level by change of variables
\cite{GR95P,BR95PP,pmr95}),
it is useful to review this work, because it bears on the nature
of the transition, in that the $B_{eff}$ from Walecka theory does give
approximately the increase in the hadronic vacuum energy necessary to make a
second-order phase transition.

As constructed by Theis et al. \cite{TG83}, the transition at baryon chemical
potential equal to zero is one from massive nucleons and antinucleons making
a transition to massless ones. These authors find a transition with zero
latent heat for the minimum coupling
\be
(g_{\sigma NN})_{min} \cong 10.8
\ee
necessary to effect a phase transition. (We consider this essentially the same
as the 
$g_{\sigma NN} \simeq 10$ of the linear sigma-omega model; the authors of
\cite{TG83} changed somewhat the parameters of the linear sigma-omega model.)
Because of the zero latent heat, Theis et al. call the transition with 
$(g_{\sigma NN})_{min}$ a second-order one. Since $M_N^*$ changes from a
small, but finite value to zero going through the phase transition, and
$M_N^*$ (or $m_Q^*$) is our order parameter, this transition would be
(perhaps weakly) first-order in our classification.

The nucleon effective mass in Walecka theory does not go to zero at finite
temperature. This is because Walecka theory includes
only positive energy nucleon (quark) states, and the nature of the
phase transition of the theory will be changed once
negative energy states are taken duly into account.

\section{The Gross-Neveu model}

\setcounter{equation}{0}

Koci\'c and Kogut \cite{KK95,KK95P} have discussed and made lattice
calculations with the higher dimensional Gross-Neveu model. On the basis of
these, they suggested mean-field critical exponents, not only for this model
but as a possibility for QCD. The works of Verbaarschot \cite{V95P} and
Jackson and Verbaarschot \cite{JV95P} support these points.

The Lagrangian of the Gross-Neveu model is
\be
{\cal L} = \bar \psi (i \partial\!\!\!/ + g\sigma) \psi - \frac{1}{2}
m_\sigma^2 \sigma^2 \;
\ee
which we shall consider for three spatial dimensions. 
Here $\sigma$ is an auxiliary
(scalar) field. Its vacuum expectation value can be obtained by
by setting the variation of the expectation value
of $\cal L$ to zero:
\be
\delta\ave{\cal L} / \delta \sigma = 0
\ee
from which we obtain
\be
\sigma = \frac{g}{m_\sigma^2} \ave{\bar\psi\psi} \;
\ee
ignoring quantum fluctuations in the $\psi$ field.
Eq.~(4.1) looks very much like Walecka theory at mean-field level, except
that negative-energy (quark) states are included in the $\ave{\bar\psi\psi}$
of eq.~(4.3) and one must take their kinetic energy into account. 
For our purposes, the Gross-Neveu model is essentially the 
Nambu--Jona-Lasinio theory without pions. Therefore, it has only $Z_2$
symmetry. However, the dynamical mass generation is basically the same as in
NJL.

The quark mass is
\be
m_Q = - g \sigma   \;,
\ee
at $\rho=0$, $T=0$. From eq.~(2.2) we see that $g$ can be identified
with $-g_{\sigma QQ}$, and $\sigma = f_\pi$ at $\rho=0$, $T=0$. The field
$\sigma$ is the order parameter, and at finite $T$ and/or $\rho$, $\sigma
\rightarrow \sigma^*$, so that $m_Q \rightarrow m_Q^*$. The chiral restoration
transition occurs when $\sigma^* \rightarrow 0$, or equivalently $f_\pi^* =
\sigma^* \rightarrow 0$. This will be the same point at which $m_Q^*
\rightarrow 0$. (In mean field, we do not have variation in $g_{\sigma QQ}$.) 

We can easily write down the gap equation for finite density, from eqs.~(4.3)
and (4.4):
\ba
m_Q^* =& 12 \,\frac{\ds g_{\sigma QQ}^2}{\ds m_\sigma^2} \left\{
       \ds\int_0^\Lambda \frac{\ds d^3 k}{(\ds 2\pi)^3} \, \frac{\ds m_Q^*}
       {\ds \sqrt{k^2 + {m_Q^*}^2}} \;-\;
       \ds \int_0^{k_F} \frac{\ds d^3 k}{\ds (2\pi)^3} \, \frac{\ds m_Q^*}
       {\ds \sqrt{k^2 + {m_Q^*}^2}}  \right \}  \cr & \cr
      =& 12 \,\frac{\ds g_{\sigma QQ}^2}{\ds m_\sigma^2} 
       \ds \int_{k_F}^\Lambda \frac{\ds d^3 k}{\ds (2\pi)^3} \, 
       \frac{\ds m_Q^*}{\ds \sqrt{k^2 + {m_Q^*}^2}} \;. \cr   
\ea
The first term in the curly bracket comes from quarks in negative energy
states, the second term from those in positive energy states.

If we assume a second-order finite density phase transition we can find the
transition point by dividing eq.~(4.5) by $m_Q^*$ and solving
the resulting equation for $m_Q^* = 0$. This leads to a critical Fermi momentum
of
\be
k_F^c \;=\; \sqrt{\Lambda^2 - \frac{\pi^2 m_\sigma^2}{3 g_{\sigma
QQ}^2}} \;=\; \Lambda \;\sqrt{ 1 - \sqrt{1 + \frac{m_Q^2}{\Lambda^2}} +
\frac{m_Q^2}{\Lambda^2} \ln (\frac{\Lambda}{m_Q} + \sqrt{1 +
\frac{\Lambda^2}{m_Q^2}})}  \;,
\ee
where the second equality follows from the vacuum gap equation.

Our construction here gives the simplest description of the $T=0$ phase
transition and it is not clear that this somewhat
schematic description holds in detail.
It is well known that the finite density transition depends on 
details of the interparticle interaction. 
Depending on the parameters we may have to add a vector interaction
term to the Lagrangian eq.~(4.1) in order to stabilize the non-trivial
solutions of eq.~(4.5). Thus the type of the phase transition (first order or
second order) may depend on the vector coupling. A vector term in the
Lagrangian, however, does not alter eq.~(4.5).

The extension of eq.~(4.5) to finite temperature is easily made by integrating
over all positive and negative energy quark states with inclusion of fermion
temperature weighting factors. For this, we follow the work of Bernard et
al. \cite{BM87}.

The bag constant can be written \cite{AB93} as
\be 
B \;=\; 12 \int_0^\Lambda \frac{d^3 k}{(2\pi)^3} \, \sqrt{k^2 + m_Q^2}
\;-\; \frac{3\Lambda^4}{2\pi^2} \;-6\;\int_0^\Lambda \frac{d^3 k}{(2\pi)^3} \, 
\frac{m_Q^2}{\sqrt{k^2 + m_Q^2}} \;,
\ee
where the first term on the right-hand side is the negative of the energy of
constituent quarks in negative energy states, the second term subtracts off
the massless quark energies they would have at chiral restoration, and the
final term is easily shown to be equal to $-\frac{1}{2} m_\sigma^2 \sigma^2$.
The result of eq.~(4.7) strongly depends on the constituent quark mass $m_Q$.
In order to reproduce the value given by eq.~(2.8) we need a rather large mass
of $m_Q \simeq 500\ {\MeV}$. 
Then for a cut-off $\Lambda = 560\ {\MeV}$, which roughly
leads to the correct values for $f_\pi$ and the quark condensate (we find
$f_\pi = 91\ \MeV$ and $\ave{\bar u u} = - (239\ \MeV)^3$), the bag constant is
\be
      B = 219\ \MeV/{\rm fm}^3  \;,
\ee
in good agreement with eq.~(2.8). Then from eq.~(4.6) we find that the phase
transition takes place at about 2.5 times nuclear matter density. This
is much too low a density to be trusted.

Although quark masses of $\sim 500\ \MeV$ can be found in the literature  
this value seems to be somewhat high. On the other hand we would expect
such large masses from the gap equation
\be
      m_Q \;=\; - \frac{g_{\sigma QQ}^2}{m_\sigma^2} \, 
                   \ave{\bar u u + \bar d d}  \;,
\ee
if we use
$g_{\sigma QQ} = 10/3$ and $m_\sigma = 705\ \MeV$, as argued in section 2.
Then for $\ave{\bar u u} = \ave{\bar d d} = - (239\ \MeV)^3$ we would find
$m_Q = 610\ \MeV$, even larger than the $500\ \MeV$ we needed to get
the result eq.~(4.8). 

However, if we insist on smaller quark masses, there is no chance to
reproduce the bag constant given by eq.~(2.8). If we choose, for instance,
$m_Q = m_N / 3 = 313\ \MeV$ we find $B = 74\ \MeV/{\rm fm}^3$ 
for $\Lambda = 650\ \MeV$
which is needed to reproduce the correct value of $f_\pi$. For $m_Q = 400\ \MeV$,
as suggested by a constituent quark picture of $\Delta$, $\rho$ and $\omega$,
we get $B = 135\ \MeV/{\rm fm}^3$ for $\Lambda = 600\ \MeV$.

At this point we should note that at $\rho = 0$, $T = 0$ the proper
variables are nucleons, as in the original Nambu--Jona-Lasinio paper, and
not quarks. The mass of the baryon which moves is $m_N$, not $m_Q$. Thus
\cite{B88}  
\be 
B \;=\; - E_{vac} \;=\; 
4\int_0^\Lambda \frac{d^3 k}{(2\pi)^3} \, \sqrt{k^2 + m_N^2}
\;-\; \frac{\Lambda^4}{2\pi^2} \;-2\;\int_0^\Lambda \frac{d^3 k}{(2\pi)^3} \, 
\frac{m_N^2}{\sqrt{k^2 + m_N^2}} \;,
\ee
where, again, the last term is $-\frac{1}{2} m_\sigma^2 \sigma^2$. Basically
the procedure in 
\cite{B88} was an extension of the Walecka theory to include the negative
energy condensate. Since each nucleon contains 3 quarks, we can make the 
connection between the nucleon and quark scalar densities as
\be
   \ave{\bar u u} + \ave{\bar d d} \;=\; 3 \ave{\bar N N} \;=\;
   \;-12\;\int_0^\Lambda \frac{d^3 k}{(2\pi)^3} \, 
   \frac{m_N}{\sqrt{k^2 + m_N^2}} \;.
\ee
Taking $\Lambda = 550\ \MeV$, we find
\be
   \ave{\bar u u} \;=\; (-249\ \MeV)^3   \;.
\ee
In this case the phase transition takes place at about 4.5 times nuclear matter
density.
Calculation of the bag constant gives 
\be
   B \;=\; 278\ \MeV/{\rm fm}^3   \;.
\ee
This is rather close to $\delta B$.

As a reconciliation of this $B$, the Walecka $B$ and the $\delta B$ found
by Koch and Brown \cite{KB93} we will adopt
\be
    B_{eff} \cong 250\ \MeV/{\rm fm}^3     \;.\label{Beff}
\ee
By extending Walecka theory to include negative energy nucleon states  down to 
a cut-off $\Lambda$ we have been able to understand the value of $\delta B$ 
found in lattice calculations \cite{KB93}. It has been pointed out by Koch et
al. \cite{KB87} a long time ago that Walecka theory including negative energy
nucleon states does not saturate at any finite densities. The authors repaired
this defect of the model by introducing another
term to the Lagrangian which is to
mimic higher-order processes. In the mean-field approximation, this 
corresponds to a density dependent scalar coupling constant and the effect of
the extra term vanishes for $\rho = 0$. Therefore it does not change eq.~(4.7)
for the bag constant.

\section{The Chiral Restoration Transition}

\setcounter{equation}{0}

In this section we construct the $\rho = 0$, finite-temperature transition. 
Our basic premise is that the transition
will occur as soon as the pressure $p$
can be brought to zero (from below). This is based on the observation that
within their accuracy, lattice gauge simulations \cite{B95P}
find the pressure to be zero.
Taking eq.~(2.10), and setting $p = 0$, we find
\be
37\, \frac{\pi^2}{90} \, T_c^4  = B_{eff}  \;,
\ee
where we have used the pressure from noninteracting quarks and gluons. We
discuss corrections of order $\alpha_S$ below.
Using eq.~(\ref{Beff})) we have
\be
T_c = 147\ \MeV
\ee
in good agreement with lattice results \cite{BO92,B95P}. 

Now the pressure must be the same (zero) on the hadronic side of the phase
transition. Brown, Jackson, Bethe and Pizzochero \cite{BJ93} have shown how
this can be accomplished. Although hadrons, other than the Goldstone bosons,
go massless at $T_c$, the effective number that can enter into the pressure
equation must be the same as the number of degrees of freedom in the quarks
and gluons. In \cite{BJ93} this is brought about by including excluded volume
effects in the hadron gas. The philosophy here is analogous to that
invoked for Debye theory for
phonons. For $N$ lattice points in a crystal there are 3N degrees of freedom.
The sum over phonons is truncated at $3N$ phonons, because the number of
effective degrees of freedom cannot exceed the number of fundamental ones.
Thus, there are 37 effective hadron degrees of freedom at $T_c$. Since, as
outlined earlier, the $B_{eff}$ in the hadron sector is equal to the $\delta
B$, which follows from decondensation of quarks and gluons, the pressure is
continuous (and zero) across the phase transition.

Koch et al. \cite{KS92} fit the Bethe-Salpeter wave function of the $\pi$- 
and $\rho$-mesons involving quarks and antiquarks propagated in the spatial
direction in hot QCD for $T \gsim T_c$. The quarks had dynamically generated
mass zero. The time direction in Euclidean space is periodic in a box of
width $T^{-1}$, so it is compressed with increasing $T$. Already at $T=T_c$,
dimensional reduction seems to work. A ``funny space" was introduced by 
interchanging $z$ and $t$; i.e., $z'=t$ and $t'=z$, where the primes refer
to the funny space. Thus the funny (imaginary) time is zero and $z'$ is
periodic with period $T^{-1}$. In the funny space, the wave functions are
two-dimensional in $x$ and $y$, constant in the (compressed) $z'$; i.e.,
they are two-dimensional ``slab" wave functions.

Given the two-dimensionality, only the 
components of the spin perpendicular to the
slab in the $z'$-direction survive integration over the $\pi$- and 
$\rho$- wave functions. Thus, the hyperfine splitting goes as
$\sigma_{1z}\sigma_{2z}$, and the helicity-zero $\rho$ meson and pion wave
functions become degenerate, both being lowered in energy from the 
helicity-one $\rho$-meson wave functions by the hyperfine interaction. The
effective dimensional reduction thus leads to the longitudinal component of 
the $\rho$-meson being ``spit-out" as the $\rho$ goes massless, and forming a 
doublet with the pion. This is reminiscent of the Georgi vector limit
\cite{G90} in which chiral symmetry is realized simultaneously in the
Wigner-Weyl and Nambu-Goldstone mode.\footnote{Although we favor
the Georgi vector symmetry near $T_c$ -- and our discussion
throughout this paper are made with that structure in mind, we must admit
that there is nothing
in lattice data or in real experiments so far performed that rules out that 
the phase slightly above $T_c$ is in the standard Wigner phase in which
$SU(n_f)\times SU(n_f)$ is restored. What would differentiate the two is the
``decay constant" $f_\pi$ (and $f_s$) which would be non-zero in the Georgi
vector symmetry phase and zero in the standard Wigner phase. If 
pseudoscalar ($\pi$) and scalar ($\delta$) mesons do persist above
the critical temperature, then the Georgi vector symmetry will be favored
over the Wigner symmetry.} A microscopic model of fully polarized
instantons \cite{SS95} reproduces features consistent with
the Georgi vector limit discussed here. 

Brown and Rho \cite{BR94} describe lattice gauge calculations \cite{GL87}
that show that the hadronic vector coupling $g_V$ goes to zero as 
$T \rightarrow T_c$, giving way to the colored vector gluon exchange of 
QCD. Together with $m_V^* \rightarrow 0$, where $m_V^*$ is the in-medium vector
meson mass, this provides yet another requirement for the Georgi vector limit.  

Lattice calculations \cite{KL94P} show from the scalar susceptibility that the
scalar $\sigma$-meson mass goes to zero as $T \rightarrow T_c$. From these
calculations and an (unpublished) extension, the $T=1$ scalar $\delta$-meson
mass goes to zero a few $\MeV$ above $T_c$. The work of Verbaarschot 
\cite{V95P} shows that the $\delta$-meson mass actually goes to zero at $T_c$,
once the bare quark mass is taken to zero. 

Since the $T=1$ scalar and $T=0$ scalar become degenerate at $T=T_c$, it is
reasonable to assume that $\pi$ and $\eta$ also become degenerate. 
This may mean  the
restoration of $U_A(1)$-invariance. 
Indeed, the work of Koch et al. \cite{KS92} can be interpreted as the 
Bethe-Salpeter wave functions being states of free quark and antiquark
correlated by the Amp\`ere's law magnetic interaction. Here $\alpha_S = 0.20$
was used. Thus, to order zero in $\alpha_S$, the dynamics is trivial,
i.e., 
that of free quarks, correlated in order $\alpha_S$. Since a quark can move
about from being correlated with another, the correlations being formed at
order $\alpha_S$, effectively deconfinement -- in the sense outlined --
accompanies chiral restoration (even though $\sim 50\%$ of the gluon
condensate remains at $T \sim T_c$). 

In summary, the dynamics above $T_c$ are those of free quarks, correlated
in order $\alpha_S$, and, therefore, are rather trivial. The Georgi vector
limit is that of nearly free quarks with, however, 
nontrivial dynamics. The nontrivial aspect is that the
hadronic vector coupling $g_V$ goes to zero as $T \rightarrow T_c$, whereas
the scalar $\sigma$ and pseudoscalar pion couplings continue smoothly through
$T_c$ \cite{HK94}. The continuity of the scalar and pseudoscalar degrees of
freedom was easy to understand in the Pisarski-Wilczek picture \cite{PW84}
where the $\sigma$ and $\pi$ are order parameters of the second-order phase
transition. In our interpretation of the transition as mean field, the 
order parameter is 
the chiral partner of the scalar meson, so it is not surprising
that these degrees of freedom carry on through the transition.

\section{Density and Temperature Dependent Masses}

\setcounter{equation}{0}

The medium dependence of vector meson masses is of great interest these days
because of the dilepton experiments at SPS at CERN \cite{A95,M95}. Calculations
with medium-dependent masses have been carried out \cite{LK95P} treating the
vector mesons as composed of constituent quarks and antiquarks, and then
implementing the density dependence of the constituent quark/antiquark by
Walecka theory. In this way, Brown/Rho scaling is realized as
\be
m_N^*/m_N \approx m_\rho^*/m_\rho \approx m_\omega^*/m_\omega \;.
\ee
The $a_1$ meson is included by using $m_{a_1}^* = \sqrt{2} m_\rho^*$. In
fact, with nonperturbative corrections, 
Brown-Rho scaling should be \cite{BR91,friman} 
\be
\frac{m_N^*}{m_N} = \frac{\sqrt{g_A^*}}{\sqrt{g_A}} \frac{m_\rho^*}{m_\rho} \;,
\ee
but this has not yet been included in the calculations of heavy-ion processes.

As noted earlier, the Walecka theory lets $m_Q^* \rightarrow 0$ only as 
$\rho \rightarrow \infty$. This is a consequence of the restriction to positive
energy states. The Walecka theory has the advantage that its parameters at
$\rho \sim \rho_o$ can be obtained from nuclear phenomena. In the case of the
Gross-Neveu model, which more properly describes the phase transition, the
order parameter $\sigma^*$ moves smoothly to zero. The Gross-Neveu model has
not, however, yet been employed in the relativistic transport calculations.
In order to use it, one would have to transport negative energy quarks,
describing their collisions and mean fields in detail assuming equilibrium 
during the fireball expansion.
The accuracy of equilibration was evaluated by Cassing et al. \cite{CE95}
for dilepton production using bare meson masses. Comparison with results by
Li et al. \cite{LK95P} who assumed equilibration, gave little difference in
the results. In the case of medium-dependent masses there are many more degrees
of freedom, and equilibration should be even better.  
Using the Gross-Neveu model is unlikely to make a large 
difference, however, because dileptons from $\rho$-meson effective masses
$m_\rho^* < 2 m_\pi$ will chiefly come out in the low-energy dilepton peak
dominated by the background of Dalitz pairs, and it will not be easy to
distinguish the low-energy dileptons from the latter.

Koch and Brown \cite{KB93} found that the increase in entropy in the region of
$T = T_c$ in lattice calculations could be well described by the
Nambu--Jona-Lasinio dependence
\be
\frac{m^*}{m} = \frac{{\ave{\bar\psi\psi}}^*}{\ave{\bar\psi\psi}} \;,
\ee
rather than $m^*/m$ scaling as
$({\ave{\bar\psi\psi}}^*/\ave{\bar\psi\psi})^{1/3}$ as naively
suggested by Brown and
Rho \cite{BR91}. Now following
Koci\'c and Kogut \cite{KK95P}, we propose to make the following argument. 
Define two relevant critical exponents $\beta$ and $\nu$ by
\be
\ave{\bar\psi\psi}^* \propto t^\beta \;,\qquad M^* \propto t^\nu
\ee
for small $t = |T-T_c|/T_c$, where $M^*$ is some physical scale like a
dynamically generated quark or hadron mass. Near $T_c$ one can do dimensional
reduction with $M^*$ as the only scale. Then the following relation
between the order parameter and the scale holds:
\be
\ave{\bar\psi\psi}^* \propto {M^*}^{\beta/\nu}   \;. 
\ee
The mean-field critical exponents are
\be
\beta = \frac{1}{2} \;,\qquad \nu = \frac{1}{2} 
\ee
so that $\beta/\nu = 1$. Of course, what we call Nambu--Jona-Lasinio
dependence comes from a mean-field calculation, so the result is readily
understandable.

The three-dimensional $O(4)$ magnet suggested in \cite{G91} and \cite{W92}
has $\beta/\nu = 3/2$ so that $M^* \propto (\ave{\bar\psi\psi}^*)^{2/3}$.
This dependence could not have been ruled out by Koch and Brown \cite{KB93},
but $\beta/\nu = 1$ was definitely favored. Thus, although the relation
(6.1) of Brown/Rho scaling appears to hold at mean-field level, as already
suggested by Brown and Rho \cite{BR95P}, the proportionality $m^*/m  =
(\ave{\bar\psi\psi}^*/\ave{\bar\psi\psi})^{1/3}$ is probably not right.

\section{The Softest Point}

\setcounter{equation}{0}

In our description of the chiral transition occurring when the
hadron masses become zero, and our limitation of the number of hadrons involved
to the number of underlying degrees of freedom, it became clear that the phase
transition would take place at the lowest temperature where the pressure
could be 
brought non-negative; i.e., at zero pressure. Lattice gauge simulations show
this quite accurately to be true \cite{B95P}. The energy density just above
the phase transition is, then,
\be
\varepsilon = 37\,\frac{\pi^2}{30}\,T^4 \;+\; B_{eff} \;=\; 4 B_{eff}
\label{7.1}
\ee
in the approximation of free quarks and gluons. 
For $B_{eff} \sim 250\ \MeV/\fm^3$,
this gives $\varepsilon \sim 1\ {\rm GeV}/\fm^3$ in agreement with
what was found by the MILK collaboration \cite{B95P}.

In relativistic heavy ion collisions, a finite baryon number chemical potential
must be introduced. The addition to the pressure is
\be
\delta p = T_c^2\,\mu_B^2 \;+\; \frac{\mu_B^4}{2\pi^2} \;.
\ee
The $\mu_B$ would be the baryon chemical potential for the (massless) nucleons
in the beam. Again, the phase transition will take place once the temperature
is reached such that the pressure is non-negative; i.e., at
\be
\frac{37\pi^2}{90}\,T_c^4 \;+\; \delta p(T_c) \;-\; B = 0
\ee
where $\delta p$ is the additional pressure (7.2) from the finite baryon
number. In calculations to date, substantial additional pressure is produced by
the repulsion from vector meson exchange between the baryons. 
As noted earlier, Brown
and Rho \cite{BR94} have discussed lattice calculations of the quark number
susceptibility which show that hadronic vector exchange cedes to perturbative
colored gluon exchange at $T = T_c$. We have  interpreted this to
be in accord with chiral symmetry
being realized in the Georgi vector limit \cite{G90}. 

The vanishing of the hadronic vector interaction is necessary in order to
understand \cite{K95} the ``cool" kaons observed in preliminary data of the
E814 and E877 experiments \cite{S94}. Confirmation of these experiments by
further data would give support to our scenario here. We plan to carry out
the fireball evolution quantitatively for the AGS energies.
Note that for a nearly stationary fireball to be formed, the vector mean 
field must be zero as in the Georgi vector limit. If it were not so,
there would not be enough energy  in the AGS experiments for the colliding
nuclei to get on top of each other, since the effective bag constant
energy $4B_{eff}$ uses up, all by itself, the total AGS energy.
(See eq.(7.1).)

The addition of $\delta p(T_c)$ in eq.~(7.3) will lower the temperature
$T_c$ of the phase transition as one can see
in a model with dilated quarks in heat
bath \cite{klr95}. Our previous argument still goes through that
the energy density will be $4 \delta B$. We have, however, restricted our
theoretical considerations here to $SU(2) \times SU(2)$ and experiments show
that a substantial amount of strangeness is produced. Correcting for this will
raise the energy density $\varepsilon$, somewhat above $1\ {\rm GeV}/\fm^3$.

For the central Si + A collisions at Brookhaven AGS, including the
E814 and E877 experiments, a rather complete set of hadron yields has 
been analyzed\cite{peter95}. The authors find a high degree of equilibration
in the products with a freezeout temperature of
\be
T_{fo}\sim 120\, -\, 140\ {\rm MeV}.
\ee
Not only do these small systems seem to equilibrate before freezeout, which has
surprised many workers in the field, but {\it there is no indication that
strangeness is not in chemical equilibrium.} It seems therefore highly plausible
that strange particles are nearly equilibrated.

In order to reach a high degree of equilibration, our scenario of 
dropping masses is definitely helpful \cite{BJ93} although perhaps not 
indispensable. When the negative energy
condensate is broken up in a relativistic heavy-ion collision, it
no longer acts as a coherent source of scalar meson field  to give the
hadron masses. The mass of the scalar degrees of freedom is
$m_\sigma \sim 500$ MeV, so the time of collision with exchange
of these degrees of freedom is
\be
\tau\sim \hbar/m_\sigma c\sim 0.4\ {\rm fm/c}.
\ee
In a time somewhat longer than this, 
the hadrons should go (nearly) massless independently  of equilibration.
In fact, in our scenario, the hadrons should go massless more quickly than
they equilibrate. Furthermore going massless in turn will speed up equilibration
because there is a large increase in the number of hadrons, as 
we have discussed in connection with the phase transition.

In the case of strange particles, Brown and Rho \cite{BR95P} showed that
the scalar mean fields act on the nonstrange quarks and antiquarks in them
very much as in  nonstrange hadrons. Additionally, considerable
$\langle \bar{s}s\rangle$ is produced in the heavy-ion collision,
and this brings down the mass of the strange quark \cite{asakawa}.
Thus, although not massless, the strange particles have much 
reduced masses in the fireball formed in the relativistic heavy-ion collisions,
so we can understand how they more or less equilibrate.

The analysis of Braun-Munzinger et al. \cite{peter95} uses a bag
constant of $B\approx 260\ {\rm MeV}/{\rm fm}^3$, essentially
the value we arrived at. In our theory, this bag constant is just
the scalar field energy  which raises the vacuum energy smoothly as
the hadrons become massless as $T\rightarrow T_c$. Even though
the growth in field energy is smooth, most of it takes place 
at $T$ very near $T_c$ as indicated in lattice calculations.

On the hadron side of the chiral transition, Braun-Munzinger
et al. do not bring the nucleons massless --
in fact, they are still essentially nonrelativistic at the phase transition,
going over to massless particles above $T_c$. Thus, a full analysis
carried out in our scenario, where the hadrons go massless at $T_c$,
would give somewhat different values for the various parameters. In 
particular, baryon number densities in the hadron gas will be closer
to those in the quark-gluon plasma if nucleons in both phases go
massless at $T_c$.

\section{Discussions}

\setcounter{equation}{0}

The Walecka theory is not manifestly chirally invariant, so near
the chiral phase transition, it may require non-trivial
modification to be reliable. Nonetheless, it is 
effectively chirally invariant at long-wavelength regime in the sense that
the $\omega$ and $\sigma$ degrees of freedom entering in Walecka theory
are present in chiral Lagrangians in the form of multi-Fermi interactions.
In any event, it reproduces well a large amount of nuclear phenomenology.
Thus, the parameters of the Walecka theory, when interpreted correctly, could
be close to the chiral ones even 
when it comes to giving a description of the chiral
phase transition.

We find that the Gross-Neveu model, essentially Nambu--Jona-Lasinio without
the pions, in mean-field approximation, is useful in making the description of
the chiral transition. 

Amusingly, the bag constant we need for the chiral transition is neither the
MIT bag constant nor that from the trace anomaly, but has a value $\sim 1/2$
of the latter. We find that it is just the field energy in Walecka theory
would result if $m_N^*$ were brought to zero by the scalar field. We give
a chiral description of it in the Gross-Neveu model, where the value turns out
to be about the same. In all cases, our treatment is based on the lattice
gauge calculations which show that about half of the condensate 
is melted as the temperature $T$ goes upward through
$T_c$. This $\sim 50$ \% is not only a property of the calculation with
dynamical quarks, as analyzed by Koch and Brown \cite{KB93}, but the same
$\sim 50$ \% was found for the quenched case by Adami et al. \cite{ahz91}.
Thus, the behavior found for the condensate is presumably a property of
the vacuum.

As we discussed earlier, the field energy in the Walecka linear sigma-omega
model nicely mocks up the $\sim 50$ \% of the condensate. We were able
to parameterize the Gross-Neveu model which contains the dynamical
symmetry breaking and mass generation so as to contain the same $\sim 50$ \%.
It is clear from this that we need two scales for the glue:
``hard" glue and ``soft" glue.\cite{AB93} The hard glue stays
rigid, like epoxy, through the phase transition, whereas the soft 
glue is melted. The division into these two components must be
$\sim 50/50$. We shall return to a more formal discussion of this below.

It is interesting to compare our scenario with that of the instanton
model of Shuryak and collaborators \cite{SV} which also predicts an 
$\sim 50$ \% disappearance of the condensate at 
the chiral restoration transition.
In this model of the vacuum which for $T\ll T_c$ chiefly involves the
``random instanton liquid," for $T\sim T_c$, about half of the condensate 
is in the random instanton liquid (soft glue) and about half in
``instanton-anti-instanton molecules" (hard glue): as $T$ approaches
$T_c$, the liquid changes from random to include an $\sim 50\%$ component
organized into molecules. Now the latter hard component which leaves chiral
symmetry unaffected, remains as $T\rightarrow T_c$. In fact the molecules are
highly polarized \cite{schaeferetal}. The fact that an instanton and 
anti-instanton pair can be laid around a torus in the time direction
at $T\sim T_c$, so that they just touch, makes the polarized molecule
a particularly favorable configuration, and helps to explain why $\sim
50\%$ of the system is in molecules for this temperature.

Whereas  the random instanton liquid disappears at $T=T_c$, the molecules
will be slowly squeezed out with increasing temperature as $1/T$ becomes
smaller. This squeezing out should follow the scenario of Gross, Pisarski 
and Yaffe \cite{GP81}
and the scale at which they disappear should be $\sim 3 T_c$. Thus from the
point of view of what happens at $T_c$, the glue in instanton molecules
is hard (``epoxy"). Above $T_c$ the quarks in the instanton molecules 
have the thermal energies given by the Matsubara frequencies, $\omega_n=(2n+1)
\pi T$.

From the standpoint of physics at temperature $T\lsim T_c$ (we do not go
far below $T_c$ because nothing happens there, at least for $\rho=0$), the glue
represented by the trace anomaly
\be
\theta_\mu^\mu=\frac{\beta (g)}{2g} {{\rm Tr}}\ G_{\mu\nu}G^{\mu\nu}
\ee
may, in analogy to the instanton picture with ``random instanton liquid" and
``instanton molecules" components, be split 
into a soft part and a hard part \footnote{A similar
splitting was studied by other authors \cite{miransky} in a different,
though related, context.}
\be
\theta_\mu^\mu=(\theta_\mu^\mu)_{soft} +(\theta_\mu^\mu)_{hard}.
\ee
In order to make a connection with Brown-Rho scaling which was formulated
in terms of an effective chiral Lagrangian implemented with trace anomaly, 
we should identify
\be
(\theta_\mu^\mu)_{soft}\approx (\chi^\star)^4
\ee
where $\chi^\star$ is the {\it soft} part of the effective scalar field,
to be identified with the dilaton $\sigma$ at the point of Weinberg's 
mended symmetry. Its vacuum expectation value, denoted simply as $\chi^\star$,
was used by Brown and Rho for scaling; thus
\be
f_\pi^\star=\left(\frac{\chi^\star}{\chi_0}\right)f_\pi\label{fpistar}
\ee
and masses were scaled similarly. It was assumed that a vacuum potential
$V(\chi^\star/\chi_0)$ ensured that $\chi^\star$ had the correct
expectation value. We do not know how to compute the vacuum potential from
the first principles but we conjecture that the role of the potential is played
by the bag constant in the Walecka or Gross-Neveu model; i.e., these models
involve effective masses $m^\star$ etc., so through (\ref{fpistar})
they are functions of $(\chi^\star/\chi_0)$.

The hard glue -- ``epoxy" -- in the instanton molecules does not scale 
appreciably in the region of $T\sim T_c$. On the other hand, it does not 
involve quark zero modes, so it does not participate in the chiral symmetry
breaking or restoration.

As we argued, the Walecka field energy or 
equivalently effective bag constant (\ref{Beff})
(or the Gross-Neveu $B_{eff}$, constructed to roughly equal (\ref{Beff}))
is just the increase in vacuum energy needed to bring the hadrons massless.
This is equivalent to the condensation energy of the random instanton liquid.
Because of the reorganization of some of the random liquid into molecules,
this is less than if the entire system were random liquid. In both cases,
the chiral restoration transition involves the hadrons going massless.

We do not have such a detailed picture for density effects. However, 
from the chiral Lagrangian and chiral identities, we can calculate
variations of relevant parameters with density \cite{BR95PP};
i.e.,
\be
\frac{f_\pi^\star (\rho_0)}{f_\pi}=0.78.
\ee
We expect that the same argument about effective potential 
$V (\chi^\star/\chi_0)$ holds here.

Recently the work by Verbaarschot and Jackson and Verbaarschot has been 
substantially extended \cite{WS96}\cite{JS96p}. The success of the random
drawing of chiral matrix elements to describe the QCD chiral restoration 
transition implies that the situation is sufficiently complicated that
spatial variations average out. Since the drawn quark matrix elements are
constants, independent of space and time, the description arrived at by
random drawing is clearly a mean field theory. 
Indeed, the random matrix model can be shown to correspond to
Nambu--Jona-Lasinio theory in mean field approximation with the kinetic energy
terms and Matsubara modes other than the lowest one being neglected \cite{NRZ}.

As practised, the random drawing includes many more matrix elements than the
zero modes, which enter into the random instanton liquid. The latter, in
themselves, would give an NJL mean field description by the methods of 
ref.~\cite{NRZ}. Basically, only the symmetries and the randomness are needed.
The hard component of the glue, described by instanton molecules, cannot be
obtained by random drawing, but must be put in by hand \cite{WS96}.

We argue for nearly trivial dynamics of the quarks and gluons correlated in
order $\alpha_S \sim 0.2$, once $T$ exceeds $T_c$. Our arguments are, however,
based on the mean-field description of the phase transition and a somewhat
rough interpretation of not-too-accurate lattice data, so we may well have
missed more subtle phenomena. We argue that effectively deconfinement takes
place at the chiral transition, in that quarks can move relatively freely,
although there is a premium for them to be correlated in color singlets
\cite{KS92}.

We find in relativistic heavy ion collisions that the ``softest point," that of
essentially zero pressure, should correspond to an energy density
$\varepsilon \gsim 1\ {\rm GeV}/\fm^3$, just
about what is reached in AGS collisions at
$11 - 15\, {\rm GeV}/N$.
Analysis of the CERES and HELIOS-3 dilepton experiments \cite{LK95P} shows
in some detail how Walecka theory applied with the parameters used here can
reproduce the results.

\newpage

\end{document}